\documentclass[aps,prl,10pt,twocolumn,groupedaddress]{revtex4-2}
\usepackage{graphicx}
\usepackage[caption=false,labelformat=empty]{subfig}
\usepackage{amsmath,amsfonts,amssymb}
\usepackage{dsfont}
\begin{document}

\title{Self-organized Collapse of Societies} 

\author{Alexander Jochim}
\email{ajochim@uni-bremen.de}
\affiliation{Institute for Theoretical Physics, University of Bremen, Germany}
\author{Stefan Bornholdt}
\email{bornholdt@itp.uni-bremen.de}
\affiliation{Institute for Theoretical Physics, University of Bremen, Germany}

\begin{abstract}
Why are human societies unstable? Theories based on the observation of recurring patterns in historical data indicate that economic inequality, as well as social factors are key drivers. So far, models of this phenomenon are more macroscopic in nature. However, basic mechanisms at work could be accessible to minimal mathematical models. 
Here we combine a simple mechanism for economic growth with a mechanism for the spreading of social dissatisfaction. Broad wealth distributions generated by the economic mechanism eventually trigger social unrest and the destruction of wealth, leading to an emerging pattern of boom and bust. We find that the model time scales compare well with empirical data. The model emphasizes the role of broad (power law) wealth distributions for dynamical social phenomena. 
\end{abstract}

\maketitle 

Collapse of societies and major political instabilities within societies have been recurring phenomena throughout history. Rather than being singular events, such major shifts are surprisingly frequent and exhibit characteristic patterns \cite{lee1931periodic,sorokin1937social,tainter1988collapse,goldstone1991revolution,turchin2003historical,turchin2008arise,turchin2009secular,turchin2010political,acemoglu2013nations,turchin2016ages,kondor2023explaining}. While a total societal collapse is less frequent, major outbreaks of political instability have been observed about every few centuries \cite{turchin2009secular,turchin2016ages,kondor2023explaining}.
From a dynamical systems perspective, this is an interesting observation and raises the question of what internal mechanisms in a society could lead to such an intrinsic instability. Turchin and colleagues isolated possible factors from a broad empirical analysis of recorded history and find that the distribution of wealth, elite structure and the state play significant roles \cite{turchin2009secular,turchin2016ages}.

One factor that we find particularly interesting in this analysis is the possible effect of wealth distribution on societal collapse, which will be the focus of this paper. 
The distribution of wealth is typically broad and can largely be modeled by a power law with lower bound, as pointed out by Pareto and supported since by many examples \cite{pareto1897,levy1997new,druagulescu2001exponential,krapivsky2010wealth,piketty2014inequality,benhabib2018skewed,zucman2019global,saez2020rise,blanchet2022generalized}.
Note that power laws challenge our intuition, for example when we try to estimate the mean wealth in such a distribution: It is dominated by either the lower end of the distribution, or by the upper end, depending on the exponent of the power law. One may say that, in the first case, most of the assets are located with the majority of the agents, while in the other case, most assets are located with the fewer richest agents. 
We here hypothesize, that this might affect the perceived inequality of how wealth is distributed. We will propose a minimal model that introduces a simple feedback between the mean of a wealth distribution and the actions of agents in a model society in order to demonstrate a possible mechanism of self-organized collapse. 

Let us first note that the mean $\langle w \rangle_\infty$ of an unbounded power law diverges for exponents $\gamma \leq 2$, while it remains 0 for larger exponents: 
\begin{align}
	\langle w \rangle_\infty = 
	\begin{cases}
        \infty, & \gamma \leq 2 \\
        0, & \gamma > 2.
    \end{cases}
\end{align}
However, for a distribution with a finite number of samples, the mean must remain finite as well. By introducing a lower and upper cutoff $w_\text{min}$ and $w_\text{max}$, let us imitate the behavior of a finite sample. The mean then is 
\begin{align}
	\langle w \rangle = 
	\frac{1 - \gamma}{2 - \gamma}  \frac{w_\text{max}^{2 - \gamma} - w_\text{min}^{2 - \gamma}}{w_\text{max}^{1 - \gamma} - w_\text{min}^{1 - \gamma}}.
    \label{eq:mean} 
\end{align} 
\begin{figure}[htbp] 
\includegraphics[width=1\linewidth]{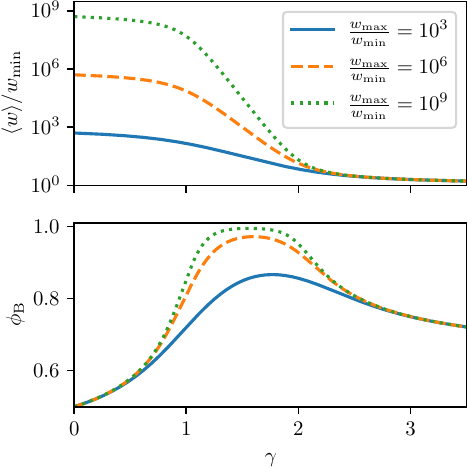} 
\caption{The mean in equation \eqref{eq:mean} in units of $w_\mathrm{min}$ (top) and the fraction of agents below mean wealth $\phi_\mathrm{B}$ (bottom) of a power law distribution with different widths $w_\text{max} / w_\text{min}$. For a uniform distribution at $\gamma=0$, exactly half of the population is below the mean. Counterintuitively, $\phi_\mathrm{B}$ is having a maximum close to $\phi_\mathrm{B} = 1$ in a range of exponents $\gamma \in (1, 2)$, and a limit of $\phi_\mathrm{B} = 1 - 1/e \approx 0.63$ for large exponents $\gamma \rightarrow \infty$.} 
\label{fig:mean_and_frac_below} 
\end{figure}

As shown in Fig.\ \ref{fig:mean_and_frac_below}, the mean of a power law with cutoffs does not diverge when lowering the exponent $\gamma$ towards $\gamma=2$. It rather transitions away from near the lower cutoff $w_\mathrm{min}$ towards near the upper cutoff $w_\mathrm{max}$ when lowering the exponent $\gamma$. However, $\gamma=2$ remains an important threshold for qualitative changes in the distribution \cite{oshanin2011proportionate}. Another counterintuitive behavior is found for the fraction of agents below the mean wealth $\phi_\mathrm{B} = \int_{w_\text{min}}^{\langle w \rangle} \mathrm{d} w \thinspace p(w)$, which exhibits a maximum in the range $\gamma \in (1, 2)$. Note that this quantity could be interpreted as the potentially unsatisfied fraction of agents. When this fraction becomes large, this could result in a larger risk for social dissatisfaction - a hypothesis which we will play out in our minimal model below. 

But first, how do such broad distributions emerge in the economic context? A number of models have been proposed  \cite{newman2005power,yakovenko2009colloquium,boghosian2014kinetics,chatterjee2015socio,jones2015pareto,greenberg2024twenty}, where the perhaps simplest mechanism is multiplicative growth with resets: In every time step, the wealth $w$ of a single agent is either multiplied by a factor $\mu$ (growth of an investment) or is reset to 1 with a probability $q$ (risk of an investment). For an ensemble of agents, this process leads to a power law probability density function $p(w) \sim w^{-\gamma}$ with constant exponent $\gamma$ as a result of the compound interest effect and risk of resets. We will use this mechanism in the model. 

\begin{figure}[htbp]
\includegraphics[width=1\linewidth]{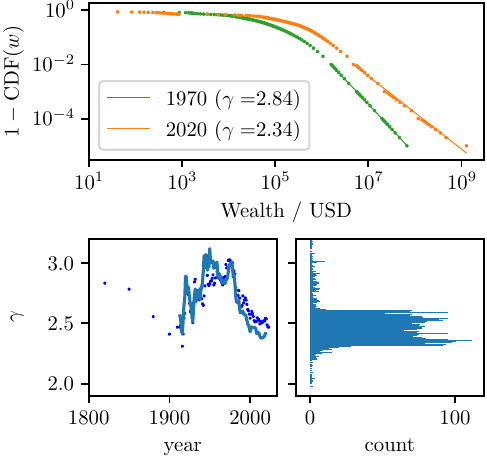}
\caption{Wealth power law tails for the United States for different years (top). Upwards and downwards trends are observed in a historical time series of the tail exponent (bottom left). The solid curve represents wealth data, and the dots income which closely resembles the exponent for wealth \cite{gaillard2023inequality}. For all countries and years available, mainly consisting of the recent decades, a histogram of exponents $\gamma$ shows an accumulation of values (bottom right). Fits were done for the top 1\% of the population on cumulative data available. For the top and bottom right plots, fine grained data of household wealth with equal split adults of age older than 20 were used. Data was taken from the World Inequality Database \cite{piketty_wid}.}
\label{fig:data}
\end{figure}

Finally, a notable aspect of empirical wealth distributions is that they are changing along history. As an example, note the US distribution of wealth and income as given in Fig.\ \ref{fig:data}.  
In our model we will likewise consider a wealth distribution that changes its slope in time. 

\begin{figure}[htbp]
\includegraphics[width=1\linewidth]{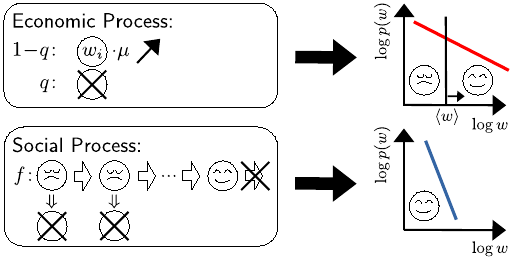}
\caption{Basic dynamics of the model. In the economic step, the wealth  $w_i$ of a random agent $i$ is either multiplied by $\mu$ (with probability $1-q$) or reset to  $w_i=1$ (with probability $q$). When repeated, this model economic process results in a power law distribution $p(w)$ where, over time, the  fraction of agents below mean wealth $\langle w \rangle$ increases due to a flattening exponent of $p(w)$. In the social step, a percolating chain of dissatisfaction is triggered with a small probability $f$. An agent $j$ compares its own wealth to the mean wealth. If it is below mean wealth, the agent resets a random neighbor $j'$ and a further new random agent $j$ repeats the process. The chain continues until an agent $j$ finds itself above mean wealth. As a result, many such resets create a steeper distribution with fewer agents below mean wealth.} 
\label{fig:model_scheme} 
\end{figure} 
Let us now define a minimal model for the two aspects: (i) economic growth that creates a wealth distribution, and (ii) a social reaction of agents who are dissatisfied with their position in the wealth distribution. Consider a finite population of $N$ agents with each agent $i$ having wealth $w_i \geq 1$. Initially, all agents start with $w_i = 1$. The model dynamics is then defined by the two following steps:

\begin{enumerate}
\item[(i)\thinspace] \emph{Economic process}: Choose a random agent $i$ and grow its wealth by $w_{i} \rightarrow w_{i} \mu $, with $\mu > 1$. With probability $q$ its wealth is reset to $w_{i} = 1$, symbolizing economic risk.    
\item[(ii)\thinspace] \emph{Social process}: With probability $f$ a chain of dissatisfaction is triggered:
    \begin{enumerate}
    \item A random agent $j$ compares its own wealth $w_{j}$ to the mean wealth $\langle w \rangle$. 
    \item If $w_{j} < \langle w \rangle$, agent $j$ is considered unsatisfied and, in an anarchic act, randomly destroys some wealth: agent $j$ picks a random other agent $j'$ and resets its wealth to $w_{j'} = 1$. After that, agent $j$ triggers another random agent: continue at step (a).
    \item Else, if agent $j$ finds its wealth above mean wealth $w_{j} \geq \langle w \rangle$, continue at (i). 
    \end{enumerate} 
\end{enumerate} 
Chains are limited to a maximum of $N$ iterations as part of the algorithm. For the simplest implementation, but not affecting the dynamics qualitatively, agents may be chosen multiple times and may also reset themselves in a chain. The mean wealth is used as a real time value at every comparison. 

A schematic description of the model steps is shown in Fig.\ \ref{fig:model_scheme}. The two consecutive model steps (i) and (ii) define one micro time step. For simulating the model, a full sweep of $N$ micro time steps defines one time unit $t \rightarrow t + 1$. Every agent $i$ is only chosen once and in different random order for every time unit.

We consider this a minimal model \cite{marsili2024simplicity} that combines the two core processes which we motivate as follows. 
In the first step, we choose the simplest form of economic growth with agents growing their wealth in proportion to their current wealth, in a multiplicative process, resembling the dynamics of an investment, with occasional resets, representing the risk of the investment. Additive processes that are responsible for the lower exponential bulk in a wealth distribution as in Fig.\ \ref{fig:data} are neglected \cite{druagulescu2001exponential}. 
 
The second step in the model is motivated by the observation that larger levels of economic inequality correlate with negative effects on society such as crime, riots, and risks of civil wars \cite{alesina1996income,wilkinson2009income,baten2013does,jerico2016does,bircan2017violent,peterson2017economic}. The underlying social mechanism draws on the phenomenon that humans tend to compare their wealth to that of others and consider it unfair if one has less. This behavioral trait has also been observed in non-human primates, where inequitable conditions elicit strong negative responses \cite{brosnan2003monkeys}. While wealth comparison is only one among several potential sources of social discontent, in the model it serves as a trigger for agent resets—interpreted as extreme responses to perceived injustice. Turchin has argued that such acts of violence can trigger chains of more violence that propagate through society in a manner analogous to cascades in epidemics, earthquakes, or forest fires \cite{turchin2016ages}. In our model, these cascades are initiated with a small probability $f$ and may percolate through the system as series of resets. Note that the key effect of the trigger chains depends on the fraction of agents with wealth below the mean: If it is large, the trigger pulse percolates through a long chain of unsatisfied agents. However if the fraction of agents below mean wealth is low, triggers do not have larger consequences. 
 
\begin{figure}
    \centering
    \includegraphics[width=1\linewidth]{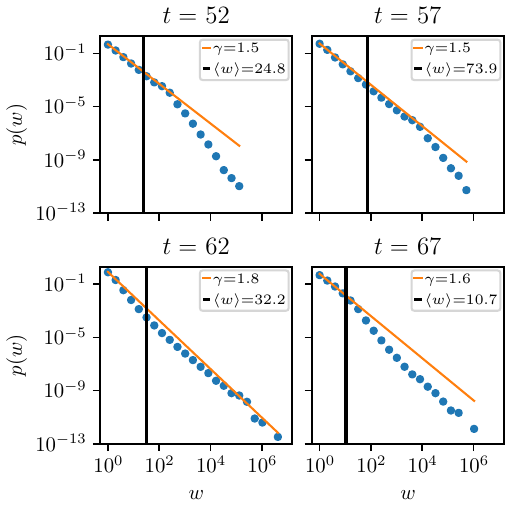}
    \caption{Example distributions for a simulation with $N=10^6$ agents, $\mu=2$, $q=0.3$ and $f=0.01$. The power law distribution shows a boom and bust behavior with a growing total wealth (boom), followed by a rapid increase of resets and decreasing total wealth, resulting in a steeper distribution (bust). Note that the total wealth is proportional to the mean wealth.}
    \label{fig:example_dist}
\end{figure}

\begin{figure}
    \centering
    \includegraphics[width=1\linewidth]{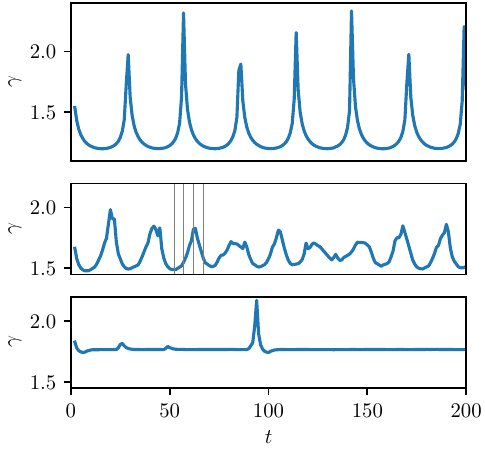}
    \caption{Time evolution of the estimated exponent $\gamma$ for three simulations with $N=10^6$ agents, $\mu=2$, $f=0.01$ and $q=0.1$ (top), $q=0.3$ (middle), $q=0.5$ (bottom). Depending on model parameters, a shift from a steady state power law to a boom and bust behavior is found. The intermediate example corresponds to the simulation in Fig.\ \ref{fig:example_dist} with gray vertical lines marking the chosen time points.}
    \label{fig:example_time}
\end{figure}

For an example simulation see Fig.\ \ref{fig:example_dist}. 
The feedback of the mean wealth and social mechanisms results in a dynamically changing behavior of the power law distribution of wealth with the fraction of the less wealthy agents $\phi_\mathrm{B}$ acting as a controlling force of the dynamics. As the economic process constantly drives the system to broader distributions with a larger mean wealth, the fraction $\phi_\mathrm{B}$ of agents below mean wealth also grows. Consequently, chains triggered by the social process get larger, and eventually, when also the larger agents have been reset, the wealth distribution becomes much less broad. Depending on model parameters, a range of behaviors can be observed, from periodic crashes, via a quasi periodic boom and bust regime, to a relatively steady state power law regime as shown by Fig.\ \ref{fig:example_time}. Furthermore, fluctuations of large agents that run away from the rest can be observed for a few time units. Those larger levels of wealth can result in single agents correlating with the dynamics of the whole system as they can keep the mean high, even though they statistically behave the exact same way as smaller agents with the same growth rate and risk. For estimating $\gamma$ in Fig.\ \ref{fig:example_time}, we used the discrete data estimator from \cite{clauset2009power} with a minimum of $w=1$. Note that the power law fit sometimes approximates only the left region of a distribution when there is a crossover between two regions with a steeper exponent on the right. For convenience the estimator value will still be called the exponent $\gamma$ in Figs.\ \ref{fig:example_dist} and \ref{fig:example_time}, however, quantitative values of $\gamma$ should be taken with caution.

For multiplicative growth with resets, as used in the model, an analytical derivation of $\gamma$ is shown in \cite{zanette2020fat}. Here, the economic process is creating a power law with regard to $\mu$ and $q$ which results in a minimal value of
\begin{align}
    \gamma_\mathrm{min} = 1 + \log_{\mu}\left(\frac{1}{1 - q}\right). 
    \label{eq:gamma_multiplicative}
\end{align}
Other model steps can only increase the total amount of resets $q_\mathrm{tot} \geq q$, resulting in a steeper power law with larger $\gamma(f>0) \geq \gamma(f=0)$. The system is constantly driven to smaller $\gamma$ until the social process dominates. In the limit $f \rightarrow 0$, the model simplifies to multiplicative growth with resets.

A typical width of the distribution can be estimated by calculating the value the largest agent reaches with probability $ 1 / 2 = 1-(1-(1-q)^{\tau_0})^N$, resulting in
\begin{align}
    \frac{w_\mathrm{max}}{w_\mathrm{min}} \approx \left( 1 - \sqrt[\leftroot{-3}\uproot{3}N]{1/2} \right) ^\frac{1}{1 - \gamma}.
\end{align}
An agent needs $\tau_0$ time units without being reset to reach the upper end. $\tau_0$ can also be used to estimate a time scale on how fast changes of the system are taking effect
\begin{align}
    \tau_0 \approx \frac{1}{1 - \gamma} \log_{\mu} \left(1 - \sqrt[\leftroot{-3}\uproot{3}N]{1/2} \right),
\end{align}
which robustly yields values of a few centuries for $\gamma \in (2,3)$, assuming a (yearly) growth rate of a few percent and $N \in (10^6, 10^9)$, corresponding to time scales observed in empirical data. 

For $f \ll 1 / \tau_0$ a steady state power law appears as the economic process dominates the dynamics. In the other extreme, for $f \gg 1 / \tau_0$ the distribution repeatedly breaks apart. Triggering events immediately cause the distribution to collapse with the social process dominating. For $f \approx 1 / \tau_0$, chains of resets are appearing on the same time scale as agents have time to pass through the system. Note that chains of resets with different magnitudes have to be possible, in order to create a boom and bust behavior. 

Despite its similarity to some models of self-organized criticality (SOC), our model is not self-organized critical. Note that the fraction $p = \phi_\mathrm{B}$ of agents with wealth lower that mean wealth could be considered as a proxy for the exponent $\gamma$ as an order parameter of the transition, which then drives the feedback mechanism of the model society. This is reminiscent of self-organized criticality in networks where a local proxy for the order parameter is used for a similar feedback mechanism \cite{bornholdt2000topological}. However, different from SOC mechanisms, the feedback in this system leads to quasi-periodic dynamics for a range of parameters, mimicking the statistics of historical records of typical collapse time scales.

The reset chains of the social process in the model corresponds to a one dimensional subcritical percolation process with spreading probability $p = \phi_\mathrm{B}$. Depending on the distribution of wealth, trigger events can have vanishing consequences or reach larger parts of the system, as $\phi_\mathrm{B}$ gets closer to the critical percolation probability $p_c = 1$. The probability $\phi_\mathrm{B}^{d}$ for a chain of size $d$ effectively amplifies the models' reaction to the exponent change of the wealth distribution, as illustrated by the curvature of $\phi_\mathrm{B}$ in the region $\gamma \approx 2$ at the bottom of Fig.\ \ref{fig:mean_and_frac_below}. The more $\phi_\mathrm{B}$ approaches $1$ for smaller $\gamma$, the larger the resulting cascade of resets turns out to be. Amplification through percolation is a mechanism most likely at work in real societies, in particular in social media with applications as viral marketing or propaganda \cite{proykova2002social}. 

Percolation in more dimensions or on complex network architectures might be of interest for extending the current model. Also, extention of the model to adaptive networks or architectures could focus on structural dynamics as social hierarchies, as motivated by recently discussed evolution and overproduction of elites \cite{mosca1896,pareto1897,turchin2009secular,turchin2016ages}. The current model does not yet include any specific social structure and could perhaps be seen as an annealed version of such specific structural models.

For this minimal model, we focused on the wealth aspect of society. The simple coupling between economic growth and social reactions can produce a variety of collapse patterns, which are robust to variations of the model. As the mean wealth is showing a non-trivial behavior, which challenges our intuition, it might contribute to our understanding of the processes that make societies unstable.

\bibliography{literature.bib}
\end{document}